\documentclass[prd,
aps,
amsmath,
amssymb,
twocolumn,
nofootinbib,
superscriptaddress
]{revtex4-2}
\usepackage{graphicx}         
\usepackage{url}
\usepackage{ragged2e}       
\usepackage{amssymb,amsmath,xcolor}
\usepackage[caption=false]{subfig}
\usepackage[makeroom]{cancel}
\usepackage{hyperref} 
\usepackage{subfig}
\usepackage{stackengine}
\begin{document}
\title{Nonlinear Compton scattering in time-dependent electric fields: LCFA and beyond}
\author{E.~G. Gelfer}\email{egelfer@gmail.com}
\affiliation{ELI Beamlines, Institute of Physics of the ASCR, v.v.i., Dolni Brezany, Czech Republic}
\author{A.~M.~Fedotov}\email{am\_fedotov@mail.ru}
\affiliation{National Research Nuclear University MEPhI, Moscow, 115409,  Russia}
\author{A.~A. Mironov}\email{mironov.hep@gmail.com}
\affiliation{LULI, Sorbonne Universit\'e, CNRS, CEA, \'Ecole Polytechnique, Institut Polytechnique de Paris, F-75252 Paris, France}
\affiliation{Prokhorov General Physics Institute of the Russian Academy of Sciences, Moscow, 119991, Russia}
\author{S.~Weber}
\affiliation{ELI Beamlines, Institute of Physics of the ASCR, v.v.i., Dolni Brezany, Czech Republic}

\begin{abstract}
Locally constant crossed field approximation (LCFA) is a powerful tool for theoretical and numerical studies of various strong field quantum electrodynamical effects. We explore this approximation in detail for photon emission by a spinless particle in a strong time-dependent electric field. This kind of electromagnetic fields is of particular interest, because, in contrast to the comprehensively studied case of a plane wave, they are not crossed. We develop an approach for calculating photon emission probability in a generic time-dependent electric field, establish the range of applicability of LCFA, and calculate the corrections to it.
\end{abstract}

\maketitle
\section{Introduction}
Compton scattering has been already studied for almost 100 years since its discovery in 1923 \cite{compton1923}. However, the interest in theoretical \cite{raicher_pra2016,hu_pra2015,king_prd2016, dipiazza_pra2017, dipiazza_pra2018, raicher_pra2019, ilderton_pra2019, mackenroth_pre2019, raicher_prr2020, acosta_prd2020, king_pra2020, heinzl_pra2020, seipt_pra2020, dipiazza_pra2021, king_prd2021,dipiazza_prd2021,chen_prd2022}, numerical \cite{blackburn_pop2018, niel_pre2018, dipiazza_pra2019, li_prappl2019, ilderton_plb2020, delgaudio_jpp2020, venkatesh_pra2020, kampfer_pra2021} and experimental \cite{bula_prl1996, bamber_prd1999, cole_prx2018, poder_prx2018,abramowicz1, abramowicz2, meuren2020} research of this process has been lately significantly increased (see also the recent reviews \cite{fedotov_arxiv2022,gonoskov_arxiv2022}). The reason is the appearance of a new generation of powerful laser facilities, such as Astra Gemini \cite{astragemini} in the UK, Hercules \cite{yanovsky_optexpr2008} in USA, Apollon \cite{papadopoulos_optlett2017} in France, ELI Beamlines \cite{elibl,weber_mre2017} in Czech Republic, ELI NP \cite{elinp,tanaka_mre2020} in Romania, CoRELS \cite{yoon_optexpr2019} in Korea, SULF \cite{gan_optexpr2017} in China, PEARL \cite{pearl} in Russia and others (see the recent review of the existing and forthcoming petawatt-class lasers in Ref.~\cite{danson_hplse2019}).

These lasers are capable of producing strong fields with dimensionless strength 
\begin{equation}\label{a0gg1}
a_0=\frac{eE_0}{m\omega}\gg1,
\end{equation}
where $-e$, $m$ are the electron charge and mass, $E_0$ and $\omega$ are the electric field amplitude and frequency (we use such units that the speed of light $c$ and Planck constant $\hbar$ are equal to unity). In such a field an electron absorbs many photons while it radiates, this regime is called the nonlinear Compton scattering (NCS).

Upon the condition \eqref{a0gg1} NCS can be used as a bright source of gamma-rays \cite{nerush_pop2014, bashinov_epjst2014, rykovanov_jpb2014, geddes_nuclinstrb2015, gu_commphys2018, zhu_apl2018, valialshchikov_prl2021}, which have a great number of potential applications, in particular in nuclear spectroscopy \cite{albert_prstab2011,geddes_nuclinstrb2015} and medicine \cite{weeks_medphys1997, carroll_amjr2003}. A fundamental physics effect based on NCS are QED cascades \cite{bell_prl2008, fedotov_prl2010, nerush_prl2011, elkina_prstab2011, king_pra2013, bashmakov_pop2014, mironov_pla2014, gelfer_pra2015, jirka_pre2016, grismayer_pre2017, vranic_scirep2018, slade_njp2019, samsonov_scirep2019, magnusson_pra2019}, which are avalanche-like processes developing when hard photons, recurrently emitted due to NCS, are in turn converted into electron-positron pairs in the same strong field.

The usual Klein-Nishina description of the Compton effect \cite{kn1929} is valid only if the electron interacts with a single photon from the external field, that is if $a_0\ll 1$ \cite{ritus1985}. In the nonlinear regime \eqref{a0gg1} one has to take into account the interaction with an arbitrary number of photons. This can be accomplished analytically only if a complete set of analytical solutions to the Dirac equation in the external field is derivable \cite{furry_physrev1951}.

This approach has two issues. First, the Dirac equation admits analytical solutions only in a few particular oversimplified cases of external electromagnetic fields, such as e.g., the Coulomb, constant uniform and plane wave fields \cite{landau4,bagrov2014,volkov1935}, while the field configurations encountered in realistic laser-matter interactions are much more intricate. Second, even if such solutions can be constructed, the photon emission probability is represented by multiple-fold integrals over them, which are very challenging for calculations. 

Fortunately, if the field is as strong as in \eqref{a0gg1} (formulation of more precise conditions is one of the goals of this paper), then both issues can be overcome by using the locally constant crossed field approximation (LCFA) \cite{ritus1985,nikishov_jetp1964}. The essence of this approximation is that 
(i) the formation scale of NCS $\sim 1/a_0$ is so small that an external field can be considered as constant over it;
(ii) the electron in such a strong external field is typically ultrarelativistic, so that it perceives the field in a proper reference frame as almost crossed  ($|E^2-B^2|\ll E^2$, $|\mathbf{E}\mathbf{B}|\ll E^2$).
Due to (i) and (ii), the emission probability in an arbitrary strong field is about the same as in a constant crossed field (CCF) \cite{ritus1985,nikishov_jetp1964}, which depends on the Lorentz invariant quantum parameters
\begin{equation}
\begin{split}
&\chi=\frac{e\sqrt{-(F_{\mu\nu} P^\nu)^2}}{m^3},\quad  \chi'=\frac{e\sqrt{-(F_{\mu\nu} P'^\nu)^2}}{m^3},\\
&\varkappa=\frac{e\sqrt{-(F_{\mu\nu} k^\nu)^2}}{m^3},
\end{split}
\end{equation}
containing the values of the electromagnetic tensor $F^{\mu\nu}(x)$ at a local position of the electron. Here $P^\mu$ and $P'^\mu$ are the kinetic 4-momenta of the electron before and after the emission and $k^\mu$ is the 4-momentum of the emitted photon.

Photon emission probability in a CCF \cite{ritus1985,nikishov_jetp1964} is quite handy and is implemented in the numerical codes, e.g. in EPOCH \cite{brady_ppcf2010}, OSIRIS \cite{osiris} and SMILEI \cite{derouillat_cpc2018} (see also the review \cite{gonoskov_pre2015} on the numerical implementation of strong field QED processes). However, this approach relies on LCFA, which has a limited range of applicability. For example, it always fails for soft photon emission (small $\varkappa$'s), as well in the wings of  laser pulses (where $a_0\ll1$). Therefore, it is very important to determine the limits of applicability of LCFA precisely, as well as the corrections to LCFA when approaching or even going beyond these limits. 

LCFA has been proven to be reliable in a strong field limit for some specific external field configurations, notably for a monochromatic plane wave \cite{ritus1985, nikishov_jetp1964} and magnetic fields \cite{baier_jetp1968}. It was also investigated for certain fields close to a plane wave, such as focused laser beams \cite{dipiazza_pra2017,dipiazza_pra2021}. However, the list of potentially interesting field configurations is much wider. In particular, colliding laser pulses produce standing-wave-like fields that are more favorable for QED cascades onset than a single pulse \cite{bell_prl2008, fedotov_prl2010, bashmakov_pop2014, gelfer_pra2015, grismayer_pre2017, magnusson_pra2019}. In an electric field antinode of the standing wave the magnetic field vanishes, and a purely time-dependent electric field serves a model capable for quantitative estimates \cite{fedotov_prl2010, elkina_prstab2011, mironov_pla2014}.

The corrections to LCFA in a plane wave were considered in \cite{baier_jetp1981,dipiazza_pra2019,ilderton_pra2019} (see also \cite{khokonov_prl2002} for an alternative approach based on accounting for variation of the curvature of the classical electron  trajectory), and in \cite{chen_prl1988} they were calculated in a pure magnetic field. Furthermore, since the corrections were found to diverge in the limit $\varkappa\to 0$, \cite{dipiazza_pra2019,ilderton_pra2019} suggested the extensions of LCFA for plane-wave-like fields to restore a reasonable agreement with the exact result in the whole range of the emitted photon energy. Besides, a recently developed locally monochromatic approximation \cite{heinzl_pra2020}, which works well for sufficiently long but not necessarily strong plane wave pulses, was shown to reduce to LCFA in a strong field limit.

However, it is worth emphasizing that, as of now, a rigorous derivation of LCFA in a general setting is still missing. Moreover, sometimes the validity of LCFA for strong fields is even doubted in general or questioned for particular configurations of external field, see e.g. \cite{raicher_pra2016,raicher_prr2020}. This is of both fundamental and practical importance, the latter because of the mentioned extensive implementation of LCFA in the modern numerical codes. 

Here we investigate NCS in a uniform time-dependent electric field with a focus on the validity of LCFA. The particular case of such configuration, a uniformly rotating electric field, was considered in Refs. \cite{raicher_pra2016, raicher_pra2019}, see also \cite{king_prd2016,hu_pra2015}. Note, however, that the NCS probability in \cite{raicher_pra2019} was considered for a particular initial condition only, which restricted the analysis in this case. On the contrary, our goal is the calculation of the NCS probability in a generic time-dependent electric field and for arbitrary initial conditions. This allows to establish a range of applicability for LCFA, as well as to calculate the corrections to it. Note that, unlike \cite{baier_jetp1981}, our approach resolves the corrections to LCFA over the emission angles. 

For the sake of simplicity and clarity, here we restrict our consideration to a field, which is periodic in time and is confined to a plane (the uniformly rotating electric field is a particular example of such a configuration). Also, to avoid technical complications from the Dirac gamma matrices algebra, we present our method for scalar QED. Generalization to standard spinor QED is straightforward and will be reported elsewhere. 

We start with a general consideration of the NCS in scalar QED in Section II. Since the Klein--Gordon equation cannot be solved exactly in an arbitrary external time-dependent electric field, we focus on WKB solutions and show their validity for the field strength of typical interest for this process. Next, we derive the LCFA probability distribution for photon emission by a scalar particle in Section III and establish its conditions of applicability in Section IV. Section V includes the discussion of the corrections to LCFA and testing our analytical results against the numerical calculations for the particular case of a uniformly rotating electric field. The total emission probability is considered in  Section VI, and the conclusion is given in Section VII. Technical details are collected in Appendix A, and the derived analytical expression for the second-order correction to LCFA is presented in Appendix B.

\section{Photon emission by a scalar particle in a time-dependent electric field}
Let us consider a spinless `electron' represented by a scalar field $\Phi$ interacting with an electromagnetic field. Following \cite{furry_physrev1951}, we split the total field into $A^{\text{tot}}_\mu=\mathcal{A}_{\mu}+A_\mu$, where  $\mathcal{A}_\mu(x)$ is the quantized radiation field (representing photons) and $A_\mu(\omega t)$ is a purely time-dependent background, assumed periodic with the frequency $\omega$. Scalar QED in such a background is governed by the Lagrangian \cite{peskin}
\begin{equation*}
	\mathcal{L} = (D^\mu\Phi)^+(D_\mu\Phi)-m^2\Phi^+\Phi-\frac{1}{4} F^2_{\mu\nu}-\frac{1}{4} \mathcal{F}_{\mu\nu}^2,
\end{equation*}
where $D_\mu=\partial_\mu-ie A_\mu^{\text{tot}}$ is the gauge covariant derivative, $\mathcal{F}_{\mu\nu}=\partial_\mu \mathcal A_\nu-\partial_\nu \mathcal{A}_\mu$ and $F_{\mu\nu}=\partial_\mu A_\nu-\partial_\nu A_\mu$ are the respective field strength tensors. The interaction with the radiation field is represented by the part 
\begin{equation}\label{Lint}
\mathcal{L}_{\mathrm{int}}=ie\mathcal{A}^\mu(\Phi^+\nabla_\mu\Phi-(\nabla_\mu\Phi)^+\Phi)+e^2 \mathcal{A}_\mu\mathcal{A}^\mu\Phi^+\Phi,
\end{equation}
where $\nabla_\mu=\partial_\mu-ieA_\mu$ is the part of the gauge covariant derivative including the external field only.
The photon and scalar fields are quantized in the Furry picture
\begin{align}\label{aop}
\mathcal{A}_\mu(x)=\int\frac{d\mathbf{k}}{(2\pi)^{3/2}\sqrt{2k}}\left(c_\mathbf{k}\epsilon_\mu e^{-ikx}+c_\mathbf{k}^+\epsilon_\mu^* e^{ikx}\right),\\
\label{fop}
\Phi(x)=\int\frac{d\mathbf{p}}{(2\pi)^{3/2}}\left(a_\mathbf{p}\Phi_\mathbf{p}+b_\mathbf{p}^+\Phi_\mathbf{p}^*\right),
\end{align}
where $k^\mu=\{k,\mathbf{k}\}$ and $\epsilon^\mu$ are the photon 4-momentum and polarization, respectively, $k=|\mathbf{k}|$; $\mathbf{p}$ is the scalar particle generalized momentum. By $c_\mathbf{k}$ and $c_\mathbf{k}^+$ we denote the annihilation and creation operators for photons, while $a_\mathbf{p}$, $a_\mathbf{p}^+$ and $b_\mathbf{p}$, $b_\mathbf{p}^+$ stand for such operators for scalar particles and antiparticles, respectively.
The scalar field modes  $\Phi_\mathbf{p}$ in Eq.~\eqref{fop} are a complete set of solutions to the Klein--Gordon equation in the external field $A^\mu$
\begin{equation}\label{KFG0}
\left(\nabla^\mu\nabla_\mu+m^2\right)\Phi_\mathbf{p}(x)=0.
\end{equation}

From now on, let us pass to the dimensionless time variable $t$ without changing the notation, $\omega t\rightarrow t$. As we assume the background field is uniform, the gauge is fixed by $A^\mu(t)=\{0,\mathbf{A}(t)\}$, so that by substitution $\Phi_\mathbf{p}(x)=e^{i\mathbf{pr}}\phi_\mathbf{p}(t)$ Eq.~(\ref{KFG0}) is reduced to
\begin{equation}\label{KFG}
\ddot{\phi}_\mathbf{p}(t)+\frac{\mathcal{E}^2(t)}{\omega^2}\phi_\mathbf{p}(t)=0,
\end{equation}
where 
\begin{equation}
\label{EP}
\mathcal{E}(t)=\sqrt{\mathbf{P}^2(t)+m^2},\quad \mathbf{P}(t)=\mathbf{p}-e\mathbf{A}(t),
\end{equation}
represent the energy and the kinetic momentum of the scalar particle, respectively.

As Eq.\eqref{KFG}  cannot be solved exactly in a general background, we proceed further by applying the WKB approximation. The WKB solutions read
\begin{equation}\label{WKB}
\phi^{(\pm)}_{\mathbf{p},\mathrm{WKB}}\approx\frac{C}{\sqrt{2\mathcal{E}(t)}}e^{\mp \frac{i}{\omega}\int\limits_{-\infty}^t\mathcal{E}(t') dt'}.
\end{equation}
As it is well known, this approximation is valid as long as $\mathcal{E}^2\gg |\dot{\mathcal{E}}|$, which in our case gives $\mathcal{E}^3\gg e|\mathbf{EP}|$, where $\mathbf{E}=-\omega\dot{\mathbf{A}}$ is the electric field. Taking into account that $\mathcal{E}\gtrsim m,|\mathbf{P}|$, one concludes that the approximation \eqref{WKB} is justified for arbitrary $\mathbf{p}$ if 
\begin{equation}\label{WKBcond}
E\ll E_\mathrm{cr}\equiv \frac{m^2}{e},
\end{equation}
where $E_\mathrm{cr}\approx 1.32\cdot 10^{16}$ V/cm is the QED critical field \cite{sauter1931,schwinger1951}. This implies that the external field should not be so strong as to induce pair production from the vacuum. It is fair to state that in the context of all practical applications of NCS this restriction is so weak that it can be always taken for granted.

Following \cite{nikishov_jetp1964,ritus1985}, we normalize the wavefunction \eqref{WKB} of the incoming particles to a particles density $n$, so that the normalization constant $C=\sqrt{n}$ for an incoming wavefunction and $C=1$ for an outgoing one.

The scattering matrix element for the NCS to the leading order reads
\begin{equation}\label{mel}
iT=\left<0\right| a_{\mathbf{p}'} c_{\mathbf{k}}\, i\int \mathcal{L}_{\mathrm{int}} d^4x\, a_{\mathbf{p}}^+ \left|0 \right>,
\end{equation}
where $\mathbf{p}$ and $\mathbf{p}'$ are the generalized momenta of the scalar particle before and after the photon emission.

Substituting Eqs.~(\ref{Lint}), (\ref{aop}), (\ref{fop}), and \eqref{WKB} into Eq.~(\ref{mel}) and integrating over $d^3x$, we obtain
\begin{equation}\label{iT}
T=(2\pi)^3e\sqrt{n}\mathcal{T}\delta(\mathbf{p}-\mathbf{p'}-\mathbf{k}),
\end{equation}
where 
\begin{equation}
\begin{split}\label{TT}
\mathcal{T}=\int\limits_{-\infty}^\infty &e^{\frac{i}{\omega}\left[ kt-\int\limits_{-\infty}^t\mathcal{E}(t')dt'+\int\limits_{-\infty}^t\mathcal{E}'(t')dt'\right]}\\
&\times \epsilon_\mu^*\frac{P^\mu(t)+P'^\mu(t)}{2\sqrt{\mathcal{E}(t)\mathcal{E}'(t)}}\frac{dt}{\sqrt{2k}}
\end{split}
\end{equation}
is the process amplitude; $P^\mu(t)=\{\mathcal{E}(t),\mathbf{P}(t)\}$, $P^{\prime\mu}(t)=\{\mathcal{E}'(t),\mathbf{P}'(t)\}$, and the primed quantities differ from Eq.~\eqref{EP} by replacing $\mathbf{p}\to\mathbf{p'}$ therein.

We can rearrange Eq.~\eqref{TT} by taking into account the periodicity of $\mathcal{E}(t)$ and $\mathbf{P}(t)$ in time.
By introducing the time-averaged energy
\begin{equation}
\mathcal{K}_{\mathbf{p}}=\frac{1}{2\pi}\int\limits_{0}^{2\pi}\mathcal{E}(t')dt',
\end{equation}
the exponent in the integrand in \eqref{TT} is cast into the following form:
\[
\begin{split}
\frac{i}{\omega}(k & -\mathcal{K}_{\mathbf{p}}+\mathcal{K}_{\mathbf{p}'})t \\
& + \frac{i}{\omega}\left[-\int\limits_{-\infty}^t\mathcal{E}(t')dt' + \mathcal{K}_{\mathbf{p}}t +\int\limits_{-\infty}^t\mathcal{E}'(t')dt'- \mathcal{K}_{\mathbf{p}'}t\right],
\end{split}
\]
where the term in the square brackets is periodic. Then the total emission amplitude $\mathcal{T}$ is rewritten as a sum of partial emission amplitudes $M_s$, each corresponding to an absorption of $s$ photons from the external field:  
\begin{equation}\label{mcT}
\mathcal{T}=\frac{2\pi}{\omega}\sum\limits_s M_s^\mu\epsilon^*_\mu\delta\left(\frac{k}{\omega}+\frac{\mathcal{K}_{\mathbf{p}'}}{\omega}-\frac{\mathcal{K}_{\mathbf{p}}}{\omega}-s\right),
\end{equation}
where the periodic part of Eq.~\eqref{TT} was expanded into a Fourier series. Here,
\begin{equation}\label{Ms}
M_s^\mu=\frac{1}{2\pi\sqrt{2k}}\int\limits_0^{2\pi} dt h^\mu(t)e^{f(t)}, 
\end{equation}
and
\begin{equation}\label{fh}
\begin{split}
&f(t)=\frac{i}{\omega}\left[kt+\int\limits_0^t\left(\mathcal{E}'(t')-\mathcal{E}(t')\right)dt'\right],\\
&h^\mu(t)=\frac{P^\mu(t)}{\sqrt{\mathcal{E}(t)\mathcal{E}'(t)}},
\end{split}
\end{equation}
the latter employs the relation $\epsilon_\mu(P^\mu+P'^\mu)=2\epsilon_\mu P^\mu$ \cite{raicher_pra2019}. 

The total photon emission rate is obtained by integration and summation of the modulus-squared matrix element (\ref{mel}) over the final states: $$W=\sum\limits_{\boldsymbol{\epsilon}}\int\frac{d\mathbf{p}'}{(2\pi)^3}\frac{d\mathbf{k}}{(2\pi)^3}|T|^2.$$ 
Since $k_\mu M_s^\mu=0$ on-shell, summation over polarization states of the emitted photon is done by the usual substitution $\sum\limits_{\boldsymbol{\epsilon}}\epsilon_\mu\epsilon_\nu^*\to -\eta_{\mu\nu}$, where $\eta_{\mu\nu}$ is the Minkowski metric tensor \cite{peskin}. Thus we obtain
\begin{equation}\label{Wsc}
\begin{split}
&\frac{1}{V_4}\frac{dW}{d\mathbf{k}}=\frac{e^2n}{4\pi^2\omega}\mathcal{R}(\mathbf{k}),\\
\mathcal{R}(\mathbf{k})=&-\sum\limits_{s}|M_s|^2\delta\left(\frac{k}{\omega}+\frac{\mathcal{K}_{\mathbf{p}'}}{\omega}-\frac{\mathcal{K}_{\mathbf{p}}}{\omega}-s\right),
\end{split}
\end{equation}
where $V_4$ is the space-time interaction volume. 

Note that for $k,\mathcal{K}_\mathbf{p},\mathcal{K}_{\mathbf{p}'}\gg \omega$ the summation over $s$ in Eq.~(\ref{Wsc}) can be replaced by integration, which has an effect of removing the $\delta$-function:
\begin{equation}\label{Mk}
\mathcal{R}(\mathbf{k})\approx -M_{\tilde{s}}^\mu M_{\tilde{s},\mu}^*,\quad \tilde{s}=\frac{k+\mathcal{K}_{\mathbf{p}'}-\mathcal{K}_{\mathbf{p}}}{\omega}.
\end{equation}
Obviously, $\tilde{s}\gg1$ has the meaning of the number of `photons' absorbed from the external field.

\section{LCFA in a time-dependent electric field}\label{sec3}

In what follows, we focus on the evaluation of $M_s$ and $\mathcal{R}(\mathbf{k})$. As we discuss below, for a strong field the integrand in Eq.~(\ref{Ms}) is a rapidly oscillating function. Therefore, following the studies of the plane-wave case \cite{nikishov_jetp1964,ritus1985}, we calculate the integral over time using the stationary phase approximation (SPA).

Let us solve the equation 
\begin{equation}\label{ens}
\dot{f}(t_0)\propto\mathcal{E}'(t_0)+k-\mathcal{E}(t_0)=0
\end{equation}
for a (complex) stationary point $t_0$ (obviously, Eq.~\eqref{ens} looks like an energy conservation but not involving the external field).

To that end it is convenient to decompose $\mathbf{P}=\mathbf{P}_\bot+\mathbf{P}_\parallel$, $\mathbf{E}=\mathbf{E}_\bot+\mathbf{E}_\parallel$, where the subscripts $\bot$ and $\parallel$ refer to the components that are transverse and parallel to $\mathbf{k}$, respectively. With this notation Eq.~\eqref{ens}  explicitly reads
\begin{equation}
\begin{split}
&\sqrt{m^2+(P_\parallel(t_0)-k)^2+\mathbf{P}_\bot^2(t_0)}+k\\
&\quad\quad\quad\quad\quad\quad-\sqrt{m^2+P_\parallel^2(t_0)+\mathbf{P}_\bot^2(t_0)}=0,
\end{split}
\end{equation}
which simplifies to
\begin{equation}\label{arben}
\mathbf{P}_\bot^2(t_0)=-m^2.
\end{equation}

Let $t_0=t_1+it_2$ ($t_{1,2}$ are the real and imaginary parts of $t_0$, respectively). Then, assuming $t_2\ll 1$ (which is justified \textit{a posteriori}), we expand 
\begin{equation}\label{p_expand}
\mathbf{P}_\bot(t_0)\approx\mathbf{P}_\bot(t_1)+ie\mathbf{E}_\bot(t_1)t_2/\omega-e\dot{\mathbf{E}}_\bot(t_1)t_2^2/2\omega,
\end{equation}
and solve Eq. (\ref{arben}) iteratively by substituting this expansion. At the leading order, the real part of the Eq.~(\ref{arben}) gives
\begin{equation}\label{t2}
t_2=\frac{\sqrt{\sigma}}{a_\bot}, \quad \sigma=1+\frac{P^2_\bot(t_1)}{m^2},
\end{equation}
where 
\begin{equation}
\boldsymbol{a}_\bot=\frac{e\mathbf{E}_\bot(t_1)}{m\omega},
\end{equation}
and $t_1$ can be determined from the imaginary part of the Eq. (\ref{arben}),
\begin{equation}\label{eqt1}
\mathbf{P}_\bot(t_1)\mathbf{E}_\bot(t_1)\approx\frac{t_2^2}{2}\frac{e}{\omega}\mathbf{E}_\bot(t_1)\dot{\mathbf{E}}_\bot(t_1).
\end{equation}

To evaluate the integral in (\ref{Ms}) we first shift the integration limits (see the details in Appendix~\ref{apa}) to get
\begin{equation}\label{Ms1}
M_s^\mu=\frac{1}{2\pi\sqrt{2k}}\int\limits_{-\pi}^\pi h^\mu\left(t_1+t\right)e^{f\left(t_1+t\right)}dt.
\end{equation}
Next, let us expand the exponent around the stationary point $t_0$ (assuming that both $t$ and $t_2$ are small) to the third order, 
\begin{equation}\label{f0_exp}
f(t_1+t)\approx f(t_0)+\frac{\ddot{f}(t_0)}2 (t-it_2)^2 +\frac{\dddot{f}(t_0)}6 (t-it_2)^3. 
\end{equation}
Up to the leading order in $a_\bot$, we have 
\begin{equation}\label{f2f3}
\ddot{f}(t_0)\approx f_2=-\frac{a_\bot^2\sqrt{\sigma}\varkappa}{\chi\chi'},\quad \dddot{f}(t_0)\approx f_3= i\frac{a_\bot^3\varkappa}{\chi\chi'},
\end{equation}
so that (cf. \eqref{t2}) $t_2=-if_2/f_3$. With account for this, we can rearrange the expansion \eqref{f0_exp} with the same accuracy as follows:
\begin{equation}\label{f0t}
f(t_1+t)\approx f^{(0)}(t)= f(t_1)-\frac{f_2^2t}{2f_3}+\frac{f_3t^3}{6}.
\end{equation}
The reason we have to keep three terms in the expansions \eqref{f0_exp}, \eqref{f0t} is that $|f_3/f_2|\sim a_\bot\gg1$, so that the second and third terms are of the same order.  As we show below, $f^{(\mathrm{IV})}(t_0)$ and higher order derivatives are at most of the order of $\propto a_\bot^3$, hence can be omitted in a leading order calculation. 

Note that in deriving Eqs.~\eqref{f2f3} we assumed that the radiating scalar particle remains ultrarelativistic both before and after the photon emission, and that it radiates closely to the direction of its propagation, i.e. that $P_\parallel\gg \max\{P_\bot,m\}$, $P'_\parallel\gg \max\{P'_\bot,m\}$ (these conditions will be analyzed in the next section). It is easy to see that with such accuracy we also have
\begin{equation}\label{chi}
\chi\approx\frac{\mathcal{E}(t_1)a_\bot\omega}{m^2},\quad \chi'\approx\frac{\mathcal{E}'(t_1)a_\bot\omega}{m^2}.
\end{equation}

By adjusting the integration contour in the complex plane (see Appendix~\ref{apa}) and expanding $h^\mu(t_1+t)$ around $t_1$ to the second-order, the integral in Eq.~(\ref{Ms1}) is evaluated to
\begin{equation}\label{Ms0}
\begin{split}
M_s^\mu\approx -\frac{1}{\sqrt{2k}} & \left\{ \vphantom{\left(\frac{2\chi\chi'}{\varkappa}\right)^{\frac{1}{3}}}   \frac{1}{a_\bot^3}\frac{\chi\chi'}{\varkappa}\ddot{h}^\mu(t_1)y\mathrm{Ai}(y)\right.\\
&+\frac{i}{a_\bot^2}\left(\frac{2\chi\chi'}{\varkappa}\right)^{\frac{2}{3}}\dot{h}^\mu(t_1)\mathrm{Ai}'(y)\\
&-\left.\frac{1}{a_\bot}\left(\frac{2\chi\chi'}{\varkappa}\right)^{\frac{1}{3}}h^\mu(t_1)\mathrm{Ai}(y)\right\},
\end{split}
\end{equation}
where 
\begin{equation}\label{y}
y=\left(\frac{\varkappa}{2\chi\chi'}\right)^\frac{2}{3}\sigma,
\end{equation}
and $\mathrm{Ai}(y)$ is the Airy function \cite{vallee2010airy}. 

The resulting leading-order contribution to the squared emission amplitude [see Eq.~\eqref{Mk}] reads
\begin{equation}\label{M0}
\begin{split}
\mathcal{R}^{(0)}=& \frac{\omega^2}{2m^2 k\chi\chi'}  \left[-\left(\frac{2\chi\chi'}{\varkappa}\right)^{2/3}\mathrm{Ai}^2(y)\right.\\
&+\left.\left(\frac{2\chi\chi'}{\varkappa}\right)^{4/3}(y\mathrm{Ai}^2(y)+\mathrm{Ai}'^2(y))\right].
\end{split}
\end{equation}
As expected, it coincides with the result in a constant crossed field \cite{ritus1985}, which manifests the LCFA.

\section{Limits of applicability of the LCFA}\label{sec_limits}
To establish the limits of applicability of the LCFA, let us recap and analyze one by one the approximations we had to make in order to obtain for the squared emission amplitude the expression Eq.~(\ref{M0}). 

(i) From the very beginning, we used the WKB approximation to solve the Klein-Gordon equation in an external field. As we have seen, this is justified for subcritical electric fields, see Eq.~(\ref{WKBcond}). 

(ii) In order to evaluate the integral in Eq.~(\ref{Ms}), we expanded the exponent $f(t)$ (as well as the pre-exponential $h^\mu$) around the stationary point to the third order [see Eq.~(\ref{f0t})] and used the SPA. To justify this approximation, one has to check when the contributions from the fourth and higher orders of such expansion are negligible. 

The interval $\Delta t$ (around $t_0$), which contributes to the integral in Eq.~(\ref{Ms}), is estimated from $|f_3|(\Delta t)^3\sim 1$ as
\begin{equation}\label{deltat}
	\Delta t\sim \frac{1}{a_\bot}\left(\frac{\chi\chi'}{\varkappa}\right)^{1/3}.
\end{equation}
Therefore, by considering 
\begin{equation}\label{f4}
	f^{(\mathrm{IV})}(t_0)\approx 3i a_\bot^3\frac{\varkappa}{\chi\chi'}\left[\frac{\boldsymbol{a}_\bot\dot{\boldsymbol{a}}_\bot}{a_\bot^2}-\frac{ma_\parallel(\mathcal{E}(t_1)+\mathcal{E'}(t_1))}{\mathcal{E}(t_1)\mathcal{E}'(t_1)}\right],
\end{equation}
we conclude that the SPA applicability condition $|f^{(\mathrm{IV})}(t_0)|(\Delta t)^4\ll 1$ results in two inequalities:
\begin{equation}\label{cond}
	\begin{split}
		&\frac{1}{a_\bot}\left(\frac{2\chi\chi'}{\varkappa}\right)^{1/3}\frac{\boldsymbol{a}_\bot\dot{\boldsymbol{a}}_\bot}{a_\bot^2}\ll1,\\
		&\left(\frac{2\chi\chi'}{\varkappa}\right)^{1/3}\frac{a_\parallel}{a_\bot}\frac{m[\mathcal{E}(t_1)+\mathcal{E}'(t_1)]}{\mathcal{E}(t_1)\mathcal{E}'(t_1)}\ll1.
	\end{split}
\end{equation}

In order to get an insight into their physical meaning, let us for a moment set aside the geometrical factors $\boldsymbol{a}_\bot\dot{\boldsymbol{a}}_\bot/a_\bot^2$ and $a_\parallel/a_\bot$. Then the first condition in Eq.~\eqref{cond} simply reads 
\begin{equation}\label{xi1}
	\xi_1=\frac{1}{a_\bot}\left(\frac{2\chi\chi'}{\varkappa}\right)^{1/3}\ll 1.
\end{equation}
Notably, a similar constraint had been obtained in Ref.~\cite{ritus1985} for the plane-wave case.\footnote{The approximation used in Ref.~\cite{ritus1985} (in fact, the SPA as well) was valid for $\beta\gg 1$, whereas in our notations $\beta=a_0^3\varkappa/(8\chi\chi')$ [see Eq.~(40) in p. 512, p. 519 and Appendix B in Ref.~\cite{ritus1985}].} Since $\xi_1$ coincides with $\Delta t$ [see Eq.~(\ref{deltat})], Eq.~\eqref{xi1} simply means that the external field can be considered as locally constant when the formation scale for the integral in Eq.~(\ref{Ms}) is smaller than the field period.

The second condition in Eq.~\eqref{cond} (for now also with the geometrical factor $a_\parallel/a_\bot$ omitted) leads to the inequalities 
\begin{equation}\label{xi2}
	\begin{split}
		&\xi_2=\frac{m}{\mathcal{E}(t_1)}\left(\frac{2\chi\chi'}{\varkappa}\right)^{1/3}\ll 1, \\
		&\xi_2'=\frac{m}{\mathcal{E}'(t_1)}\left(\frac{2\chi\chi'}{\varkappa}\right)^{1/3}\ll 1,
	\end{split}
\end{equation}
which restrict (from below) the Lorentz factors of the charged particle before and after the emission. By this, one ensures that the field looks as about crossed in the reference frame of the particle during the interaction. Such a condition could not appear in the plane wave case, for which the field is precisely crossed in any reference frame.

The geometrical factors $\boldsymbol{a}_\bot\dot{\boldsymbol{a}}_\bot/a_\bot^2$ and $a_\parallel/a_\bot$ in Eq.~\eqref{cond} can be large if $\dot{a}_\bot, a_\parallel\gg a_\bot$, namely, when the photon is emitted almost in parallel to the electric field. In such case, the original validity conditions, Eqs.~(\ref{cond}), might be even stronger than Eqs. (\ref{xi1}) and (\ref{xi2}). We will explore the corresponding example in Section~\ref{sec5}. 

One may argue that the terms in square brackets of Eq. (\ref{f4}) may cancel each other in some special cases. As a result, the first correction to the phase $f^{(0)}$ might become small
even at high values of the parameters (\ref{cond}). Nevertheless, we still conclude that LCFA should fail for such cases, as the corrections to the pre-exponential factor $h^\mu$ as well as higher-order corrections to the phase are large (see the explicit calculation of the corrections below).

(iii) The derivation assumes that the particle radiates almost along it propagation direction, i.e. that $|\mathbf{P}_\bot(t_1)|\ll P_\parallel(t_1)$. Let us show that this holds automatically under the already mentioned assumptions. Indeed, since the Airy function vanishes for large argument, only the values $y\lesssim 1$ contribute to the matrix element in Eq.~(\ref{Ms0}). Therefore, we have
\begin{equation}
	\frac{P_\bot(t_1)}{m}\lesssim\left(\frac{2\chi\chi'}{\varkappa}\right)^{1/3}.
\end{equation}
By comparing this inequality with Eq.~(\ref{xi2}), we arrive at $\mathcal{E}(t_1)\gg P_\bot(t_1)$. In the ultrarelativistic case $\mathcal{E}(t_1)\gg m$, this leads to $ P_\parallel(t_1)\gg P_\bot(t_1)$.

\begin{figure*}[t]
\topinset{(a)}{\subfloat{\includegraphics[width=0.47\textwidth]{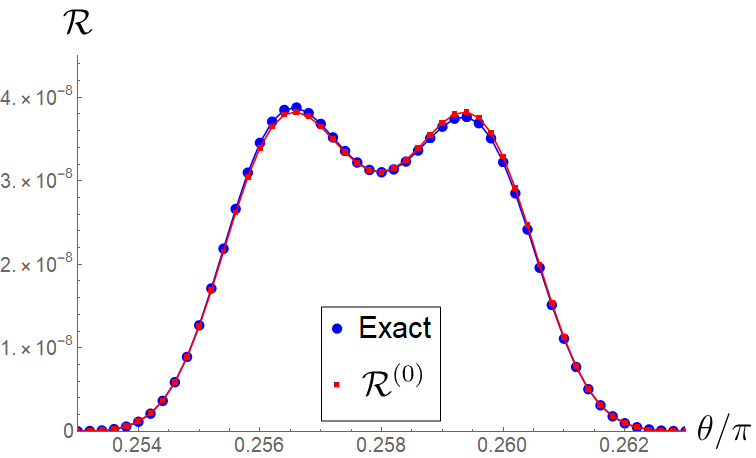}}}{0.8cm}{-2.5cm}
\topinset{(b)}{\subfloat{\includegraphics[width=0.47\textwidth]{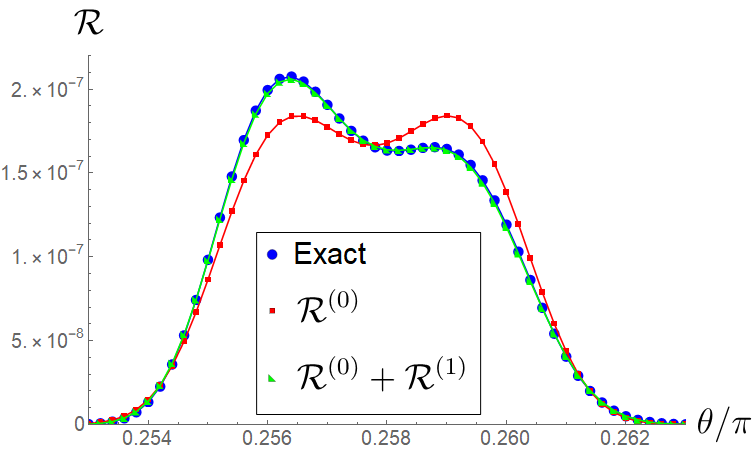}}}{1.2cm}{-2.4cm}\\
\topinset{(c)}{\subfloat{\includegraphics[width=0.47\textwidth]{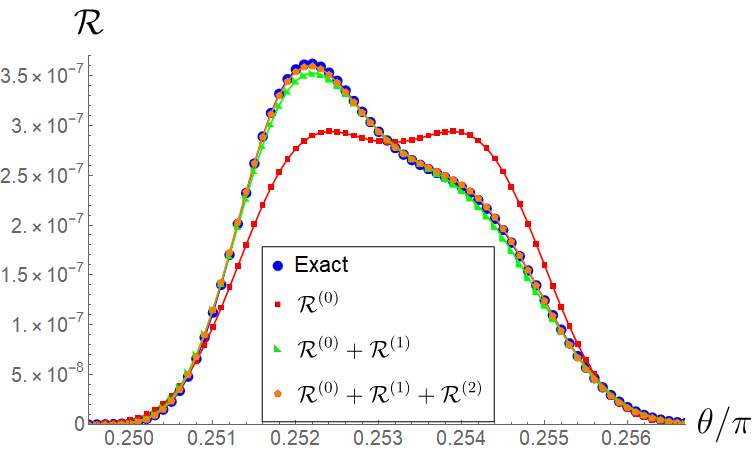}}}{1.2cm}{-2.4cm}
\topinset{(d)}{\subfloat{\includegraphics[width=0.47\textwidth]{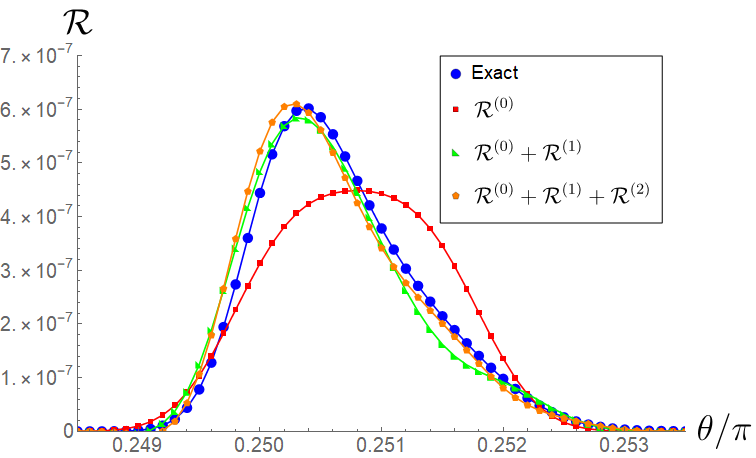}}}{1.2cm}{-2.4cm}
\caption{Squared emission amplitude $\mathcal{R}$ vs emission azimuthal angle $\theta$ for various values of the field strength: $a_0=200$ (a), $a_0=10$  (b), $a_0=3$ (c) and $a_0=1$ (d). In all plots: $\mathbf{p}=200\,m\{1,0,1\}$, $\omega=10^{-5}m$, $k=m$, emission polar angle $\varphi=0$.}\label{fig_num1}
\end{figure*}

\section{Corrections to the LCFA}\label{sec5}

Next we turn to the corrections to the LCFA differential probability rate. Specifically, let us focus on the squared photon emission amplitude \eqref{Mk}. As shown, in the leading order the amplitude is given by Eq.~\eqref{M0}. To calculate the corrections, we recast the amplitude into the form 
\begin{equation}
\begin{split}\label{Ms11}
&M_s^\mu=\frac{1}{2\pi\sqrt{2k}}\int\limits_{-\pi}^\pi h^\mu(t_1+t)e^{\tilde{f}(t)}e^{f^{(0)}(t)} dt,\\
&\tilde{f}(t)=f(t_0-it_2+t)-f^{(0)}(t),
\end{split}
\end{equation}
and expand $h(t_1+t)$ and $\exp[\tilde{f}(t)]$ into  powers of small $t$ and $t_2$, this time retaining higher orders. As in Section \ref{sec3}, we extend the integration limits to $(-\infty,\infty)$ (see also Appendix \ref{apa}), so that the resulting integrals are reduced to either Airy function or its derivatives (\ref{iairy}). This way the squared emission amplitude (\ref{Mk}) is represented as
\begin{equation}\label{Mser}
\mathcal{R}(\mathbf{k})=\mathcal{R}^{(0)}+\mathcal{R}^{(1)}+\mathcal{R}^{(2)}+\ldots,
\end{equation}
where $\mathcal{R}^{(0)}$ is the leading-order LCFA term as in Eq.~(\ref{M0}) and $\mathcal{R}^{(j\geq 1)}$ are the $j$-th order corrections in small parameters $\xi_1$, $\xi_2$ and $\xi_2'$.

\subsection{The first order correction}

When calculating the corrections, the stationary point should be also found from equation \eqref{arben} with higher accuracy. For example, for the first-order correction, one should expand $\mathbf{P}_\bot(t_0)$  up to $O(a_\bot^{-3})$. By doing so, for $t_2=\mathrm{Im}\,t_0$ we get
\begin{equation}\label{t2_1_order}
t_2\approx \frac{\sqrt{\sigma}}{a_\bot}\left[1-\frac{\boldsymbol{\tau}\dot{\boldsymbol{a}}_\bot}{2a_\bot^2}\right],
\end{equation}
where $\boldsymbol{\tau}=\mathbf{P}_\bot(t_1)/m$. Notably, $t_1=\mathrm{Re}\,t_0$ remains unaffected to this order. It is worth noting that the first-order correction to the expressions (\ref{chi}) for $\chi$ and $\chi'$ also vanishes.

The second term in Eq.~\eqref{t2_1_order} corrects $\ddot{f}(t_0)$ and $\dddot{f}(t_0)$, which now read:
\begin{equation}
	\begin{split}
		&\ddot{f}(t_0)\approx \left(1+\frac{\nu_1}{3}+\frac{\nu_2}{2}\right)f_2,\\
		&\dddot{f}(t_0)\approx \left(1+\nu_1+\nu_2\right)f_3,
	\end{split}
\end{equation}
where 
\begin{equation}\label{nu}
\begin{split}
&\nu_1=6i\left[\xi_1\left(\frac{\varkappa}{2\chi\chi'}\right)^{2/3}\frac{\boldsymbol{\tau}\boldsymbol{a}_\bot}{\sqrt{y}}-\frac{a_\parallel}{a_\bot}\frac{\sqrt{y}}{2}(\xi_2+\xi_2')\right],\\ &\nu_2=\xi_1\frac{\boldsymbol{\tau}\dot{\boldsymbol{a}}_\bot}{a_\bot}\left(\frac{\varkappa}{2\chi\chi'}\right)^{1/3}.
\end{split}
\end{equation}
 
By noting that $f^{(\mathrm{IV})}(t_0)\approx \nu_1 f_3^2/f_2$ and 
\begin{equation}
\mathrm{Re}\,\tilde{f}(0)=\mathrm{Re}\,[f(t_1)-f^{(0)}(0)]\approx \frac{\nu_2 y^{3/2}}{3},
\end{equation}
we further obtain 
\begin{equation}
\tilde{f}(t)\approx \frac{y^{3/2}}{12}\left[\nu_1(r^2-1)^2+4\nu_2r^3\right],
\end{equation}
where $r=(f_3/f_2) t=-it/(\xi_1\sqrt{y})$.

After substituting the following expansions 
\[e^{\tilde{f}(t)}\approx 1+\tilde{f}(t),\quad h^\mu(t_1+t)\approx\sum_{j=0}^3 \frac{d^j h^\mu(t_1)}{dt^j}\frac{t^j}{j!},\] 
into Eq.~\eqref{Ms11}, one is in a position to calculate the integral over $t$. By modulus-squaring the resulting expression, we finally extract the first-order correction:
\begin{equation}\label{M1}
\begin{split}
\mathcal{R}^{(1)}=&\frac{\nu_2}{3}\frac{\omega^2}{2m^2 k\chi\chi'}\left(\frac{2\chi\chi'}{\varkappa}\right)^{4/3}\\
&\times \left[ \vphantom{\left(\frac{\varkappa}{2\chi\chi'}\right)^{2/3}} 5y\mathrm{Ai}^2(y)+4y^2\mathrm{Ai}(y)\mathrm{Ai}'(y)+4\mathrm{Ai}'^2(y)-\right.\\
&\quad\quad\left.2\left(\frac{\varkappa}{2\chi\chi'}\right)^{2/3}\left(\mathrm{Ai}^2(y)+y\mathrm{Ai}(y)\mathrm{Ai}'(y)\right)\right].
\end{split}
\end{equation}
Note that $\mathcal{R}^{(1)}$ is proportional to the small parameter $\xi_1$ but does not contain $\xi_2$ and $\xi_2'$, which, however, appear in the higher order corrections, starting with $\mathcal{R}^{(2)}$.

The above-described approach can be continued to obtain systematically the corrections of any demanded order. However, the resulting expressions (not to say, the intermediate steps) turn out progressively more lengthy. For this reason we only include final expressions for the second order correction $\mathcal{R}^{(2)}$ relegating them to Appendix~\ref{apc}. Note that it is quadratic in $\xi_1$, $\xi_2$ and $\xi_2'$, thus substantiating them as a complete set of expansion parameters in the problem.
 
\subsection{Emission in a uniformly rotating electric field}

\begin{figure}[h!]
	\includegraphics[width=0.45\textwidth]{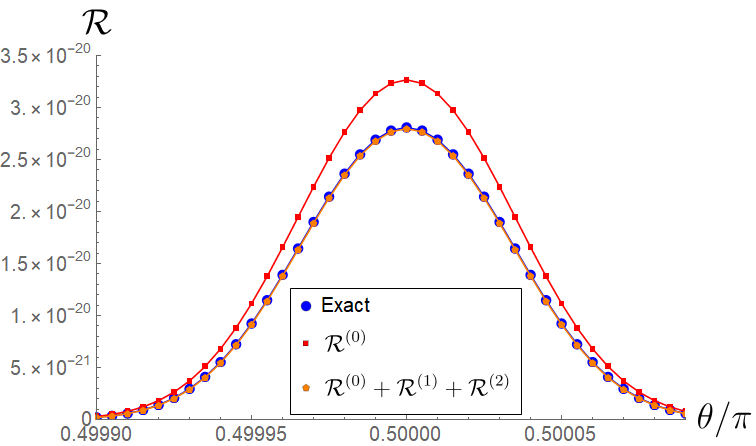}
	\caption{Squared emission amplitude $\mathcal{R}$ vs azimuthal angle $\theta$ for $a_0=2000$, $\mathbf{p}=0$, $\omega=10^{-5}m$, $k=1996\,m$, polar angle $\varphi=0$.}\label{fig_num2}
\end{figure}

\begin{figure*}[t]
	\topinset{(a)}{\subfloat{\includegraphics[width=0.47\textwidth]{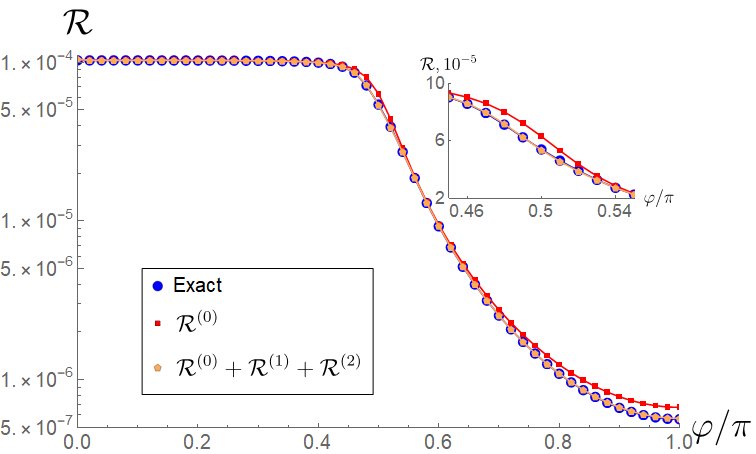}}}{1cm}{2.4cm} 
	\topinset{(b)}{\subfloat{\includegraphics[width=0.47\textwidth]{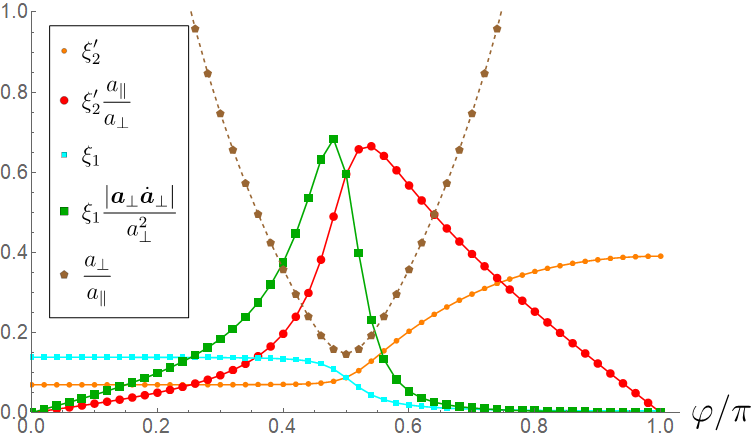}}}{1cm}{2.4cm} 
	\topinset{(c)}{\subfloat{\includegraphics[width=0.47\textwidth]{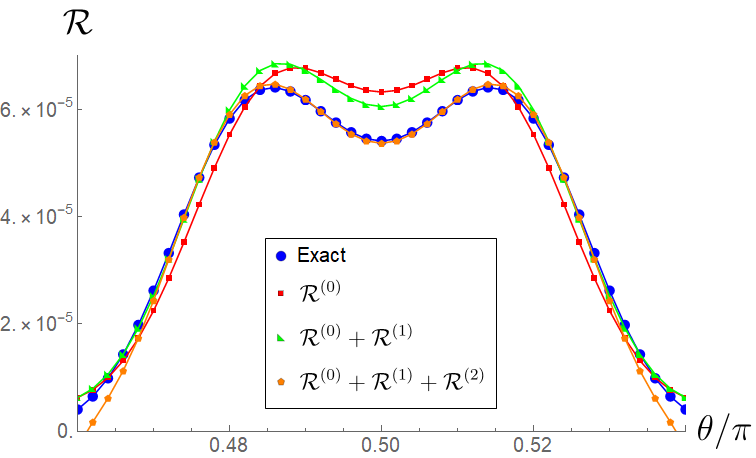}}}{1cm}{2.4cm} 
	\topinset{(d)}{\subfloat{\includegraphics[width=0.47\textwidth]{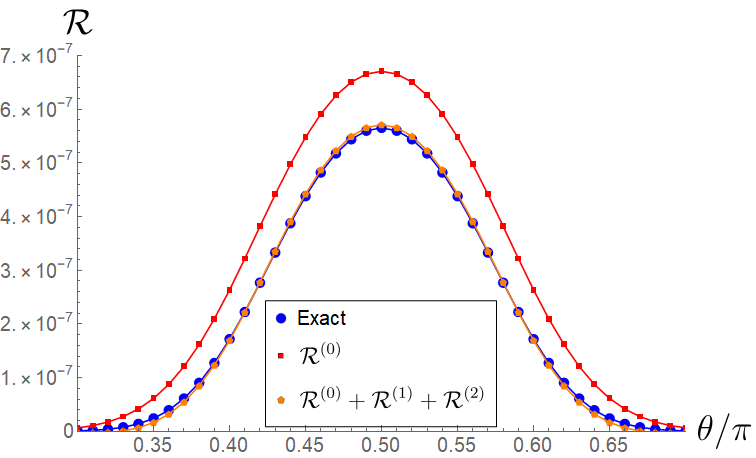}}}{1cm}{2.4cm} 
	\caption{(a) Squared emission amplitude $\mathcal{R}$ vs polar angle $\varphi$ at $\theta=\pi/2$ for the emission direction $\mathbf{k}$ [inset: the same dependence on $\varphi$ magnified in the range near $\varphi=0.5\pi$]. (b) The expansion parameters [see Eqs.~(\ref{cond}), (\ref{xi1}), (\ref{xi2})] vs $\varphi$ for fixed $\theta=\pi/2$. (c) and (d) $\mathcal{R}$ vs azimuthal angle $\theta$ at $\varphi=\pi/2$ and $\varphi=\pi$, respectively. For all the plots we set $a_0=200$, $\mathbf{p}=198\, m\{1,0,0\}$, $\omega=10^{-5}m$, $k=0.03\,m$.
	}
	\label{fig_num3}
\end{figure*}

Let us test the squared LCFA amplitude (\ref{M0}) and the corrections to it, given by (\ref{Mser}), (\ref{M1}), against the numerically evaluated initial semiclassical expression, the latter obtained by inserting Eq.~(\ref{Ms}) into Eq.~(\ref{Mk}). For brevity, we refer to the latter as the `exact' calculation.
To that end, we consider a particular case of the electric field uniformly rotating in the $xy$-plane: \[\boldsymbol{a}(t)=\frac{e\mathbf{E}(t)}{m\omega}=a_0\{\cos t,\,\sin t,\,0\},\]
and several cases of parameter selection. The results of the calculations are presented in Figs.~\ref{fig_num1}-\ref{fig_num3}.

First, we set $\mathbf{p}$ oblique with respect to the $xy$-plane (for definiteness, let it lie in the $xz$-plane) and assume $p>m a_0$. Notably, the kinetic momentum $\mathbf{P}(t)$ of the emitting particle covers a cone-like surface during the field rotation period. We consider the emission probability rate of a photon with the wave vector $\mathbf{k}=k\{\sin\theta\cos\varphi,\, \sin\theta\sin\varphi,\, \cos\theta\}$ and analyze the dependence of $\mathcal{R}$ on the azimuthal angle $\theta$ for $\varphi=0$ (i.e., assuming that $\mathbf{k}$ also lies in $xz$-plane) and $k$ is fixed. Fig.~\ref{fig_num1} provides such a dependence for different values of $a_0$. The particular choice of the parameters is given in the caption to Fig.~\ref{fig_num1}. 

It follows from the numerical calculations that parameters $\xi_{1}$ and $\xi_2$, $\xi_2'$ are almost constant in the range of each plot in Fig.~\ref{fig_num1}. Moreover, for these plots $\xi_2$ and $\xi_2'$ are small, so we actually track the value of $\xi_{1}$ only.

In the case of largest $a_0$ [$a_0=200$, see Fig.~\ref{fig_num1}(a)] the LCFA result $\mathcal{R}^{(0)}$ (almost) coincides with the exact calculation. Here, one has $\xi_1\approx 0.046$, so that the LCFA works precisely as expected. For $a_0=10$ one has $\xi_1\approx 0.26$ and $\mathcal{R}^{(0)}$ deviates from the exact calculation as seen in Fig.~\ref{fig_num1}(b). However, with account for the first-order correction $\mathcal{R}^{(1)}$ a perfect agreement between the analytical approach and the exact result is restored.

The case of $a_0=3$  [see Fig.~\ref{fig_num1}(c)] corresponds to $\xi_1\approx 0.56$. One can observe some discrepancies between the exact calculation and the first-order result. But upon including the second-order correction $\mathcal{R}^{(2)}$, the shape of the curve converges to the exact distribution. 

Finally, for $a_0=1$  (Fig.~\ref{fig_num1}(d)) $\xi_1\approx 1.17$. In this case accounting for even higher order corrections is mandatory to accurately reproduce the exact result. 

It is interesting to note that, unlike $\mathcal{R}^{(1)}$, the distribution of $\mathcal{R}^{(0)}$ in Fig.~\ref{fig_num1} is symmetric. From a physical point of view, the symmetry of the LCFA expression $\mathcal{R}^{(0)}$  with respect to $\theta$ (for a given fixed $\varphi$) can be understood as follows. The essence of LCFA is that the photon emission is formed at a small time interval centered at $t_1$, so that variation of the vectors $\mathbf{P}(t)$ and $\mathbf{E}(t)$ is negligible. Under such conditions the vectors $\mathbf{P}(t_1)$ and $\mathbf{E}(t_1)$ are the only ones specific for the problem. If so, then radiation should be symmetric with respect to the plane spanned by these vectors. 

When LCFA is violated, a broad interval of $t$ contributes to the emission amplitude \eqref{Ms}. Its central point $t_1$ may also essentially vary with respect to the angles $\theta$ and $\varphi$. Due to both reasons the  mirror symmetry may break, as is seen in the figures. We come back to this property in Sec.~\ref{apb}. 



Now let us illustrate the accuracy of LCFA with regard to the remaining parameters $\xi_2$ and $\xi_2'$. As already mentioned, they are missing in the leading-order correction (\ref{M1}), but show up in the expansion \eqref{Mser} starting from the second-order correction $\mathcal{R}^{(2)}$ onward.

Note that $\xi_2'>\xi_2$ due to energy conservation, so that smallness of $\xi_2'$ should be enough to ensure the validity of LCFA. Conversely, $\xi_2'\gtrsim1$ should be enough for the LCFA to fail. Let us consider the latter criterion in two particular cases: (i) hard photon emission, when the emitted photon carries away almost the entire  energy of the radiating particle, i.e. $\varkappa\approx\chi\gg\chi'$; (ii) emission of softer photon, for which $\varkappa\lesssim\chi'\sim\chi$. 

In case (i) we have $\chi'\ll\chi$ and the criterion $\xi_2'\gtrsim 1$ can be reformulated as
\begin{equation}\label{cfv1}
	\frac{1}{\chi'}\left(\frac{a_\bot}{a_{cr}}\right)^{3/2}\gtrsim 1,
\end{equation}
where 
\begin{equation}
	a_{cr}=\frac{m}{\omega}
\end{equation}
corresponds to $a_0$ for the critical field [see Eq.~(\ref{WKBcond})]. 

In case (ii), $\chi'\sim\chi$, large $\xi_2'$ is achieved with
\begin{equation}\label{cfv2}
	\frac{1}{\chi'}\left(\frac{a_\bot}{a_{cr}}\right)^3\gtrsim \varkappa,
\end{equation}
or equivalently
\begin{equation}\label{cfv22}
	\frac{m}{\mathcal{E}'(t_1)}\frac{a_\bot}{a_{cr}} \gtrsim \frac{k}{m}.
\end{equation}

In Fig.~\ref{fig_num2} we show an example of the first case, namely, hard photon emission violating the LCFA due to that the particle loses almost the entire energy so that the field cannot be considered as almost crossed in the rest frame of the particle after the emission. Here both parameters $\xi_1\approx\xi_2\approx 3\times 10^{-4}$ are small, but $\xi_2'\approx 0.13$ is such that the corrections to LCFA are considerable. As previously, by taking into account the corrections up to the second-order, we reinstate a good agreement with the exact calculation. 

It is worth noting that the emission probability of such a hard photon is small, see Fig.~\ref{fig_num2}. Indeed, given that Eq.~(\ref{cfv1}) holds and $\chi\sim\varkappa$, the argument of the Airy functions in $\mathcal{R}^{(j)}$ [see Eq.~\eqref{y}] can be estimated as $y\gtrsim a_{cr}/a_\bot\gg1$, meaning that the probability is exponentially suppressed. 

Next consider the emission of a softer photon corresponding to the condition in Eq.~(\ref{cfv22}), which restricts from below the energy $\mathcal{E}'\sim\mathcal{E}$ of the particle. To illustrate this condition, we choose $\mathbf{p}$ directed along $x$-axis, so that it now lies in the plane of the field. We  set $|\mathbf{p}|\approx ma_0$ and vary $\mathcal{E}$ from the values satisfying Eq.~(\ref{cfv22}) downward, so that $\mathcal{E}(t_1)\sim m$ for certain emission directions $\mathbf{k}=k\{\sin\theta\cos\varphi,\, \sin\theta\sin\varphi,\, \cos\theta\}$.

First, let us consider the photon emission in the plane $\theta=\pi/2$ of the rotating electric field, see Fig.~\ref{fig_num3}(a). If the photon is emitted along $\mathbf{p}$ (small polar angles $\varphi$), then both the particle energy $\mathcal{E}(t_1)$ and the transverse component $a_\bot(t_1)$ of the field are large. Consequently, all the expansion parameters $\xi_1$, $\xi_2$, $\xi_2'$ are small, see Fig.~\ref{fig_num3}(b) and therefore the LCFA result coincides with the exact calculation. 

Next, let us study the case when the photon is emitted transversely to $\mathbf{p}$ ($\varphi=\pi/2$). The dependence of $\mathcal{R}$ on $\theta$ is given in  Fig.~\ref{fig_num3}(c). It shows that the LCFA notably deviates from the exact values. While all the parameters $\xi_1$, $\xi_2$, $\xi_2'$ remain relatively small, the geometrical factors in Eq.~\eqref{cond}, namely, $a_\parallel/a_\bot$ and $|\boldsymbol{a}_\bot\dot{\boldsymbol{a}}_\bot|/a_\bot^2$, appear to be large [see Fig.~\ref{fig_num3}(b)], as $\mathbf{k}$ is almost parallel to the electric field $\mathbf{a}$. Note that in this case the terms containing both $\xi_1$ and $\xi_2$, $\xi_2'$ [see Eq.~\eqref{second_corr}] contribute to the correction $\mathcal{R}^{(2)}$.

Finally, assume the photon is emitted opposite to $\mathbf{p}$ ($\varphi\approx\pi$). This implies that $\mathbf{k}$ and $\mathbf{P}(t_1)$ are (almost) parallel, and since $p\approx ma_0$, one has $\mathcal{E}(t_1)\sim m$. Therefore, for certain values of $k$ the LCFA breakdown condition in Eq.~(\ref{cfv22}) can be fulfilled. We show such an example in Fig.~\ref{fig_num3}(d). Notably, for this plot one has $\mathcal{E}(t_1)\ll ma_\bot$, so that $\xi_1\ll 1$. This means that in this case LCFA is violated solely due to $\xi_2$ and $\xi_2'$ approaching 1, i.e. the field can be still considered as locally constant, albeit no longer crossed.

\section{Total emission probability}\label{apb}
 
Let us now briefly discuss the total photon emission probability, which is obtained by integrating the distribution Eq.~(\ref{Wsc}) over $d\mathbf{k}$. In the present work, we consider only the total LCFA probability  and the leading order correction to it. 

\begin{figure*}[t]
	\includegraphics[width=0.5\textwidth]{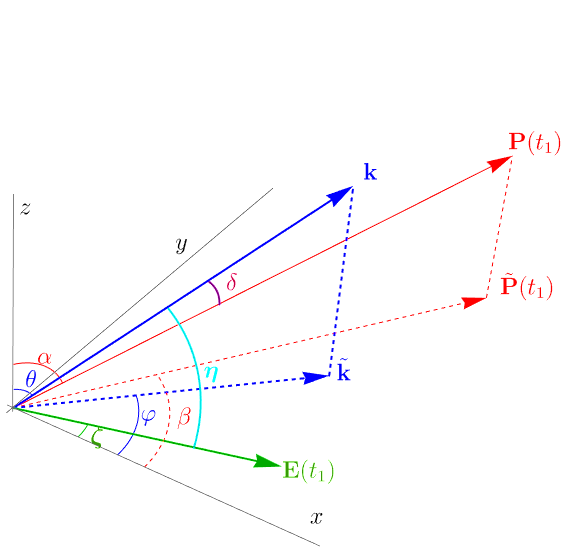}
	\caption{The electric field $\mathbf{E}$, the momentum $\mathbf{P}(t_1)$, its projection to the $(x,y)$ plane $\tilde{\mathbf{P}}$, the wave vector $\mathbf{k}$, and the angles involved.}\label{v3d}
\end{figure*}

It is convenient to change the variables $\{k_x,k_y,k_z\}\rightarrow\{u,\,\tau,\,t_1 \}$, where
\begin{equation}
	u=\frac{\varkappa}{\chi-\varkappa},
	\quad 
	\tau =\frac{|\mathbf{P}_\bot(t_1)|}{m}\mathrm{sign}(\alpha-\theta).
\end{equation}
Note that for convenience we define $\tau$ with an additional $\mathrm{sign}(\alpha-\theta)$ with respect to $|\boldsymbol{\tau}|$ defined after Eq.~\eqref{t2_1_order}. Here and in the following, we introduce the azimuthal and polar angles $(\theta,\varphi)$ and $(\alpha,\beta)$ for the vectors $\mathbf{k}$ and $\mathbf{P}(t_1)$, respectively, and the following additional notations: $\delta$ for an angle between $\mathbf{P}(t_1)$ and $\mathbf{k}$, $\eta$ --- between $\mathbf{E}(t_1)$ and $\mathbf{k}$, $\zeta$ --- between $\mathbf{E}(t_1)$ and $x$-axis, and $\mu$ --- between $\mathbf{E}_\bot(t_1)$ and $\mathbf{P}_\bot(t_1)$, see Fig.~\ref{v3d} for reference.  
These angles are related as follows:
\begin{equation}\label{angles}
	\begin{split}
		&\sin\alpha\cos(\beta-\zeta)=\cos\delta\cos\eta+\sin\delta\sin\eta\cos\mu,\\
		&\cos\delta=\cos\alpha\cos\theta+\sin\alpha\sin\theta\cos(\beta-\varphi),\\
		&\cos\eta=\cos(\varphi-\zeta)\sin\theta.
	\end{split}
\end{equation}

Expressing $\mathbf{k}$ explicitly in terms of the new variables is challenging. However, as long the expansion Eq.~\eqref{Mser} for $\mathcal{R}$ is valid, it is enough to find such expressions only to the same required order in $\xi$-parameters. For our first example, consider the leading order obtained by inserting $\mathcal{R}^{(0)}$ in Eq.~\eqref{M0} into the probability Eq.~\eqref{Wsc}. Correspondingly, we keep the leading order in the variable change.

As the particles are assumed ultrarelativistic, we have $P_\bot(t_1)\ll P_\parallel(t_1)$, $|\alpha-\theta|\ll 1$, $|\beta-\varphi|\ll 1$ and $\delta\ll 1$ and can expand Eqs.~(\ref{angles}) in small arguments. It follows from Eq.~(\ref{eqt1}) that the term
\begin{equation}
	\sin\delta\cos\mu\approx \xi_1\xi_2\frac{y}{2}\frac{\boldsymbol{a}_\bot\dot{\boldsymbol{a}}_\bot}{a_\bot^2}= O(\xi_1\xi_2),
\end{equation}
and therefore can be neglected. Thus we obtain
\begin{equation}\label{phi1}
	\beta-\varphi\approx (\alpha-\theta)\cot\alpha\cot(\beta-\zeta), 
\end{equation}
and
\begin{equation}\label{delta}
	\delta\approx|\theta-\alpha|\sqrt{1+\cos^2\alpha\cot^2(\beta-\zeta)}.
\end{equation}

With this, it is now possible to express the old variables through the new ones explicitly as follows:
\begin{equation}\label{cvar}
	\begin{split}
		&k=\frac{u\chi m^2}{(1+u)a_\bot\omega},\\
		&\theta\approx\alpha-\frac{m\tau}{P(t_1)\sqrt{1+\cos^2\alpha\cot^2(\beta-\zeta)}},\\
		&\varphi \approx\beta-\frac{m\tau\cot\alpha\cot(\beta-\zeta)}{P(t_1)\sqrt{1+\cos^2\alpha\cot^2(\beta-\zeta)}}.
	\end{split}
\end{equation}
This provides the Jacobian determinant of the transformation:
\begin{equation}\label{Jacob}
	\left|\frac{\partial(k,\theta,\varphi)}{\partial(u,\tau,t_1)}\right|\approx \frac{m^2 a_\bot}{P(t_1)\sin\alpha(1+u)^2}.
\end{equation}
By taking into account that at the leading order $\theta\approx \alpha$, we arrive at the LCFA expression \cite{nikishov_jetp1964,ritus1985}
\begin{equation}\label{W0}
	\frac{W}{V_4}\approx\frac{1}{2\pi}\int\limits_0^{2\pi}\mathcal{P}^{\mathrm{LCFA}}(t_1)dt_1,
\end{equation}
where
\begin{equation}\label{P0}
	\begin{split}
		&\mathcal{P}^{\mathrm{LCFA}}(t_1)=\frac{e^2m^2n}{2\pi P(t_1)}\int\limits_0^\infty \frac{du}{(1+u)^2} \int\limits_{-P(t_1)/m}^{P(t_1)/m} d\tau\\
		&\times\left(\frac{u}{2\chi}\right)^{\frac{1}{3}}\left[-\mathrm{Ai}^2(y)+\left(\frac{2\chi}{u}\right)^{\frac{2}{3}}\left(y\mathrm{Ai}^2(y)+\mathrm{Ai}'^2(y)\right)\right].
	\end{split}
\end{equation}
Since $P(t_1)/m\gg 1$, we can further extend the limits of the integral over $\tau$ to $\pm\infty$ and evaluate it \cite{nikishov_jetp1964,ritus1985}. Thus we obtain the spectral distribution of the emitted photon:
\begin{equation}\label{dpdu0}
	\frac{d\mathcal{P}^{\mathrm{LCFA}}}{du}=-\frac{e^2m^2 n}{4 P(t_1)}\left[\mathrm{Ai}_1(z)+\frac{2}{z}\mathrm{Ai}'(z)\right],
\end{equation}
where $z=(u/\chi)^{2/3}$ and $\mathrm{Ai}_1(z)=\int_z^\infty \mathrm{Ai}(y)dy$.

Let us now consider the first-order correction $\mathcal{R}^{(1)}$. According to Eqs.~(\ref{M1}) and (\ref{nu}) it is proportional to the scalar product 
\begin{equation}\label{angles1}
	\boldsymbol{\tau}\dot{\boldsymbol{a}}_\bot=|\boldsymbol{\tau}|\dot{a}_\bot\cos\mu_1,
\end{equation}
where $\mu_1$ is the angle between $\mathbf{P}_\bot(t_1)$ and $\dot{\mathbf{E}}_\bot(t_1)$. Let us introduce the angles $\zeta_1$ between $\dot{\mathbf{E}}(t_1)$ and the $x$-axis and $\eta_1$ between $\dot{\mathbf{E}}(t_1)$ and $\mathbf{k}$. They satisfy the relation
\begin{equation}
	\begin{split}
		&\sin\theta\cos(\varphi-\zeta_1)\cos\delta+\sin\eta_1\sin\delta\cos\mu_1=\\
		&\quad\quad\quad\quad\quad\quad\quad\quad\quad\quad\quad\quad\quad\quad\sin\alpha\cos(\beta-\zeta_1),
	\end{split}
\end{equation}
which is analogous to Eq.~(\ref{angles}). By using Eq.~\eqref{angles1} in the ultrarelativistic limit and taking into account that $|\dot{\boldsymbol{a}}_\bot|=|\dot{\boldsymbol{a}}(t_1)|\sin\eta_1$ and Eq.~(\ref{delta}), we arrive at
\begin{equation}\label{taua}
	\begin{split}
		&\boldsymbol{\tau}\dot{\boldsymbol{a}}_\bot\approx\frac{\tau|\dot{\boldsymbol{a}}_\bot(t_1)|\cos\alpha}{\sin\eta_1\sqrt{1+\cos^2\alpha\cot^2(\beta-\zeta)}}\\
		&\quad\quad\quad\times[\cos(\beta-\zeta_1)-\cot(\beta-\zeta)\sin(\beta-\zeta_1)].
	\end{split}
\end{equation}

In virtue of \eqref{nu} and \eqref{taua}, the first-order correction \eqref{M1} to the squared amplitude is odd in $\tau$. This is the reason for the asymmetric shape of the first-order correction in Fig.~\ref{fig_num1}. As the correction to the Jacobian determinant from Eq.~(\ref{Jacob}) is also odd, the first-order correction to the total probability in Eq.~(\ref{W0}) vanishes identically. 





Another particular consequence of (\ref{taua}) is that $\mathcal{R}^{(1)}$ is proportional to $\cos\alpha$, meaning that it also vanishes if the initial momentum $\mathbf{p}$ [hence also $\mathbf{P}(t_1)$] lies in the same plane as the electric field.


\section{Conclusions}

We have calculated the probability distribution for photon emission by a scalar particle in a strong time-dependent electric field, assuming the field is subcritical, periodic, and planar. The result is represented by a power series in the parameters $\xi_1$, $\xi_2$ and $\xi_2'$, defined in Eqs.~\eqref{xi1}, \eqref{xi2}. The zeroth-order term coincides with the LCFA [see Eq.~\eqref{M0}], and the corrections can be systematically calculated up to any required order. In particular, we present the first- and the second-order corrections to the LCFA modulus-squared emission amplitude, which determine the differential  distribution of photon emitted with momentum $\mathbf{k}$ [see Eqs.~\eqref{M1} and \eqref{second_corr}].

The expansion parameters Eqs.~\eqref{xi1}, \eqref{xi2}  depend on the transverse (with respect to the emission direction) component of the field strength, quantum dynamical parameters and energies of the incoming and outgoing particles and have a clear physical meaning. Namely, smallness of $\xi_1$ is equivalent to stating that the time interval, which contributes to the integral representing the matrix element [see Eq.~\eqref{Ms}], is much smaller than the period of the external field. Obviously, in such a case the field can be considered locally constant. The conditions $\xi_2\ll1$ and $\xi_2'\ll1$ mean that the particle is ultrarelativistic both before and after photon emission, so that the field appears almost as crossed in the particle reference frame. 

We have tested the LCFA and the first two corrections to it against the numerically evaluated squared emission amplitude for the case of a uniformly rotating electric field. In particular, we have investigated angular distributions at fixed photon energies. As long as the expansion parameters $\xi_1$, $\xi_2$ and $\xi_2'$ are small, the LCFA result is in perfect agreement with the exact calculation. But as they approach unity, the discrepancy in the shape of the distributions becomes visible, and accounting for the corrections is necessary. By doing so, an excellent agreement with numerical calculations can be reestablished. A significant deviation from the LCFA is observed for either very soft or very energetic (those draining almost the entire energy from the emitting particle) photons, or in emission at large angles (including backwards) with respect to the generalized momentum of the particle $\mathbf{p}$. At higher energies of the photon (yet such to maintain the outgoing particle ultrarelativistic) LCFA remains a good approximation. The latter observation is extremely important for simulation of self-sustained QED cascades \cite{elkina_prstab2011, raicher_pra2019}.

Our approach can be generalized for fermions in a straightforward way. This, along with the analysis of higher-order corrections to the photon energy spectrum and the total emission rate, will be addressed elsewhere.

\section*{Acknowledgements}
E.G.G. and S.W. were supported by the project
ADONIS (Advanced research using high intensity laser produced photons and particles)
CZ.02.1.01/0.0/0.0/16\_019/0000789
from European Regional Development
Fund. S.W. was also supported by High Field Initiative (HiFI)(CZ.02.1.01/0.0/0.0/15\_003/0000449) from European Regional Development
Fund.
A.A.M. acknowledges funding from the Russian Foundation for Basic Research (Grant No. 19-32-60084) and from  Sorbonne Universit\'e in the framework of the Initiative Physique des Infinis (IDEX SUPER). 
A.M.F. was supported by the MEPhI Program Priority 2030 and Russian Foundation for Basic Research (Grant No. 20-52-12046).

\appendix
\section{Calculation of $M_s^\mu$}\label{apa}

In order to evaluate the integral in Eq.~(\ref{Ms}), we first note that due to the presence of the $\delta$-function in Eq.~(\ref{Wsc}) the integrand is a periodic function along the horizontal lines in the complex $t$-plane since
\begin{equation}
f(t+2\pi)=f(t)+2\pi is.
\end{equation}
For this reason we can shift the integration limits to arrive at Eq.~(\ref{Ms1}).

Then, using Eq.~(\ref{f0t}), we get
\begin{equation}\label{Ms2}
M_s^\mu\approx\frac{1}{2\pi\sqrt{2k}}\int\limits_{-\pi}^\pi h^\mu(t_1+t)e^{f(t_1)-\frac{f_2^2}{2f_3}t+\frac{f_3}{6}t^3}.
\end{equation}
Note that $f(t_1)$ is imaginary, so that doesn't contribute to the mode-square, and that $f_3=i|f_3|$ (see Eq.~\eqref{f2f3}).

\begin{figure}[h!]
\includegraphics[width=0.7\linewidth]{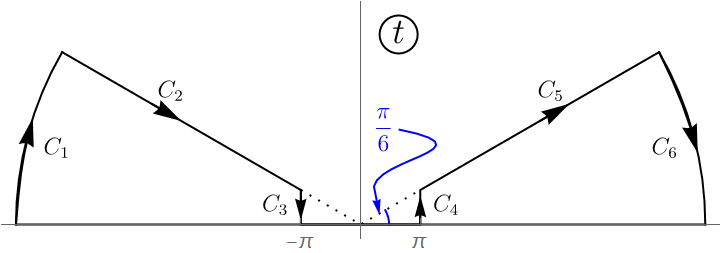}
\caption{Integration contour for evaluation of the integral in Eq.~(\ref{Ms2}). The lines $C_2$ and $C_5$ make the angles $5\pi/6$ and $\pi/6$, respectively, with the real axis.}\label{fig_contour}
\end{figure}

According to Fig.~\ref{fig_contour}, 
\begin{equation}
\begin{split}
\int\limits_{-\pi}^\pi h^\mu(t_1+t)e^{f^{(0)}(t)}dt =&\int\limits_{-\infty}^\infty h^\mu(t_1+t)e^{f^{(0)}(t)}dt \\ &+\int\limits_{C} h^\mu(t_1+t)e^{f^{(0)}(t)}dt,
\end{split}
\end{equation}
where the integration path $C=C_1+C_2+C_3+C_4+C_5+C_6$. The integrals over the remote arcs $C_1$ and $C_6$ vanish and the sum of the integrals over the vertical segments $C_3$ and $C_4$ is equal to zero because of the $2\pi$-periodicity of the integrand along the real axis. Consider 
\begin{equation}
\begin{split}
\int\limits_{C_5} h^\mu(t_1+t)e^{f^{(0)}(t)}dt=&\int\limits_{\frac{2\pi}{\sqrt{3}}}^\infty h^\mu\left(t_1+\rho e^{\frac{i\pi}{6}}\right)\\
& \quad\quad\times e^{f(t_1)-\frac{|f_3|\rho^3}{6}-\frac{f_2^2\rho}{2|f_3|}e^{-\frac{i\pi}{3}}}d\rho,
\end{split}
\end{equation}
where $t=\rho e^{i\pi/6}$. The value of the integral is formed at the lower limit and is exponentially small for $a_\bot^3\gg \chi\chi'/\varkappa$. Therefore, we neglect the contributions to $M_s^\mu$ from $C_2$ and $C_5$.

Finally, by expanding $h^\mu(t_1+t)$ in Eq.~(\ref{Ms2}) around $t_1$ up to the second-order and using
\begin{equation}\label{iairy}
\begin{split}
&\int\limits_{-\infty}^\infty t^k e^{i\frac{f_2^2}{2|f_3|}t+i\frac{|f_3|}{6}t^3}dt=2\pi (-i)^k\left(\frac{2}{|f_3|}\right)^{\frac{k+1}{3}}\frac{d^k\mathrm{Ai}(y)}{dy^k},
\end{split}
\end{equation}
where $y=\left(\frac{2}{|f_3|}\right)^{1/3}\frac{f_2^2}{2|f_3|}$ corresponds to Eq.~(\ref{y}), we obtain Eq.~(\ref{Ms0}).

\begin{widetext}
\section{Second-order correction}\label{apc}
For completeness, let us provide the resulting  expression for the second-order correction to the squared emission amplitude (a derivation goes along the same lines as described for first-order correction in the main text):
\begin{equation}\label{second_corr}
\mathcal{R}^{(2)}=\frac{\omega^2}{m^2\chi\chi'k\rho^2}\left[\xi_1^2\mathcal{R}^{(2)}_{11}+\xi_2^2\mathcal{R}^{(2)}_{22}+{\xi'_2}^2\mathcal{R}^{(2)}_{2'2'}+\xi_1\xi_2\mathcal{R}^{(2)}_{12}+\xi_1\xi'_2\mathcal{R}^{(2)}_{12'}+\xi_2\xi'_2\mathcal{R}^{(2)}_{22'}\right],
\end{equation}
\begin{equation}
\begin{split}
&\mathcal{R}^{(2)}_{11}=\frac{1}{60}\left\{y^2 \left[30 \left(\frac{\boldsymbol{a}_\bot\dot{\boldsymbol{a}}_\bot}{a_\bot^2}\right)^2-32 \frac{\boldsymbol{a}_\bot\ddot{\boldsymbol{a}}_\bot}{a_\bot^2}-39 \left(\frac{\dot{a}_\bot}{a_\bot}\right)^2\right]\text{Ai}^2(y) \right.\\
&\quad\quad\quad\quad\quad\quad+2 y \left[15 \left(\frac{\boldsymbol{a}_\bot\dot{\boldsymbol{a}}_\bot}{a_\bot^2}\right)^2-2 \left(7 \frac{\boldsymbol{a}_\bot\ddot{\boldsymbol{a}}_\bot}{a_\bot^2}+9 \left(\frac{\dot{a}_\bot}{a_\bot}\right)^2\right)\right] {\text{Ai}'}^2(y)\\
&\quad\quad\quad\quad\quad\quad+\left. \left[45 \left(\frac{\boldsymbol{a}_\bot\dot{\boldsymbol{a}}_\bot}{a_\bot^2}\right)^2-4 \left(\frac{\boldsymbol{a}_\bot\ddot{\boldsymbol{a}}_\bot}{a_\bot^2} \left(4 y^3+3\right)+3 \left(\frac{\dot{a}_\bot}{a_\bot}\right)^2 \left(y^3+2\right)\right)\right]\text{Ai}(y) \text{Ai}'(y) \right\}\\
&\quad\quad\quad\quad+\frac{\rho}{180} \left\{y \left[-135 \left(\frac{\dot{\boldsymbol{a}}_\bot\boldsymbol{a}_\bot}{a_\bot^2}\right)^2+96 \frac{\ddot{\boldsymbol{a}}_\bot\boldsymbol{a}_\bot}{a_\bot^2}+72 \left(\frac{\dot{a}_\bot}{a_\bot}\right)^2-40 \left(\frac{\boldsymbol{\tau}\dot{\boldsymbol{a}}_\bot}{a_\bot}\right)^2 \left(y^3+10\right)\right]\text{Ai}^2(y)\right.\\
&\quad\quad\quad\quad\quad\quad\quad\quad+\left.5 \left[9 \left(\frac{\dot{\boldsymbol{a}}_\bot\boldsymbol{a}_\bot}{a_\bot^2}\right)^2-8 \left(\frac{\boldsymbol{\tau}\dot{\boldsymbol{a}}_\bot}{a_\bot}\right)^2 \left(y^3+7\right)\right] {\text{Ai}'}^2(y)\right.\\
&\quad\quad\quad\quad\quad\quad\quad\quad+\left.6 y^2  \left[4 \frac{\ddot{\boldsymbol{a}}_\bot\boldsymbol{a}_\bot}{a_\bot^2}+3 \left(\frac{\dot{a}_\bot}{a_\bot}\right)^2-90 \left(\frac{\boldsymbol{\tau}\dot{\boldsymbol{a}}_\bot}{a_\bot}\right)^2\right]\text{Ai}(y) \text{Ai}'(y)\right\}\\
&\quad\quad\quad\quad+\frac{\rho^2}{9} \left(\frac{\boldsymbol{\tau}\dot{\boldsymbol{a}}_\bot}{a_\bot}\right)^2 \left[y^2 {\text{Ai}'}^2(y)+8 y \text{Ai}(y) \text{Ai}'(y)+\left(y^3+5\right) \text{Ai}^2(y)\right],
\end{split}
\end{equation}
\begin{equation}
\begin{split}
&\mathcal{R}^{(2)}_{22}=\frac{1}{10}\left\{\left[\left(\frac{a_\parallel}{a_\bot}\right)^2+1\right]\left(y{\text{Ai}'}^2(y)-y^2 \text{Ai}^2(y)\right)-\left[\left(\frac{a_\parallel}{a_\bot}\right)^2 \left(8 y^3-9\right)+8 y^3+6\right] \text{Ai}(y) \text{Ai}'(y)\right.\\
&\quad\quad\quad\quad\quad\quad\quad+\left.y\left[\left(\frac{a_\parallel}{a_\bot}\right)^2+1\right]\left(\text{Ai}^2(y)+4 \text{Ai}(y)\text{Ai}'(y)\right)\right\},
\end{split}
\end{equation}
\begin{equation}
\begin{split}
&\mathcal{R}^{(2)}_{2'2'}=\mathcal{R}^{(2)}_{22}-\left[\left(\frac{a_\parallel}{a_\bot}\right)^2-\frac{1}{2}\right]\text{Ai}(y)\text{Ai}'(y),
\end{split}
\end{equation}
\begin{equation}
\begin{split}
&\mathcal{R}^{(2)}_{12}=\frac{1}{30} \left\{ \left[3 \frac{\dot{\boldsymbol{a}}_\bot\boldsymbol{a}_\bot}{a_\bot^2} \frac{a_\parallel}{a_\bot}-\frac{\dot{a}_\parallel}{a_\bot}\right] \left(y^2 \text{Ai}^2(y)- y{\text{Ai}'}^2(y)\right)\right.\\
&\quad\quad\quad\quad\quad\quad-2 \text{Ai}(y) \text{Ai}'(y) \left[4 y^3 \left(\frac{\dot{a}_\parallel}{a_\bot}-3 \frac{\dot{\boldsymbol{a}}_\bot\boldsymbol{a}_\bot}{a_\bot^2} \frac{a_\parallel}{a_\bot}\right)+6 \frac{\dot{\boldsymbol{a}}_\bot\boldsymbol{a}_\bot}{a_\bot^2} \frac{a_\parallel}{a_\bot}+3 \frac{\dot{a}_\parallel}{a_\bot}\right]\\
&\quad\quad\quad\quad\quad\quad-\left.\rho y  \left[3 \frac{\dot{\boldsymbol{a}}_\bot\boldsymbol{a}_\bot}{a_\bot^2} \frac{a_\parallel}{a_\bot}-\frac{\dot{a}_\parallel}{a_\bot}\right] \left(4 y \text{Ai}(y)\text{Ai}'(y)+\text{Ai}^2(y)\right)\right\},
\end{split}
\end{equation}
\begin{equation}
\begin{split}
&\mathcal{R}^{(2)}_{12'}=\mathcal{R}^{(2)}_{12}+\frac{a_\parallel}{a_\bot}\text{Ai}(y) \text{Ai}'(y),
\end{split}
\end{equation}
\begin{equation}
\begin{split}
&\mathcal{R}^{(2)}_{22'}=\frac{1}{30} \left\{2 \left[\left(\frac{a_\parallel}{a_\bot}\right)^2+1\right]\left( y^2 \text{Ai}^2(y)- y {\text{Ai}'}^2(y)\right)\right.\\
&\quad\quad\quad\quad\quad\quad +\left[16 \left(\left(\frac{a_\parallel}{a_\bot}\right)^2+1\right) y^3-18 \left(\frac{a_\parallel}{a_\bot}\right)^2+27\right]\text{Ai}(y) \text{Ai}'(y)\\
&\quad\quad\quad\quad\quad\quad-\left.2\rho y\left[\left(\frac{a_\parallel}{a_\bot}\right)^2+1\right]\left(\text{Ai}^2(y)+4 y \text{Ai}(y)\text{Ai}'(y)\right)\right\},
\end{split}
\end{equation}
where $\rho=(\varkappa/2\chi\chi')^{2/3}$.
\end{widetext}


%

\end{document}